\documentclass[12pt,preprint]{aastex}

\shorttitle{Discovery of relatively hydrogen-poor giants in
$\omega$\,Cen}
\shortauthors{Hema et al.}

\begin{document}

\title{Discovery of relatively hydrogen-poor giants in the Galactic \\
globular cluster $\omega$\,Centauri }

\author{B. P. Hema and Gajendra Pandey}
\affil{Indian Institute of Astrophysics, Bengaluru, Karnataka 560034, India; 
\\ hema@iiap.res.in, pandey@iiap.res.in}

\begin{abstract}

In this letter, the results of our low-resolution
spectroscopic survey for identifying the
hydrogen-deficient (H-deficient)
stars in the red giant sample of the
globular cluster $\omega$\,Cen are reported.
Spectral analyses were carried out on the basis
of the strengths of (0,0) MgH band and the Mg\,$b$
triplet. In our sample, four giants were identified
with weak/absent MgH bands in their observed spectra
not as expected for their well determined
stellar parameters. The Mg abundances for the program stars
were determined from subordinate lines of the MgH band 
to the blue of the Mg\,$b$ triplet,
 using the spectral synthesis technique.
The derived Mg abundances for the program stars were as expected
for the red giants of $\omega$\,Cen (Norris \& Da Costa 1995), except
for the four identified candidates.
Determined Mg abundances of these four candidates are
much lower than that expected for the red giants of $\omega$\,Cen,
and are unacceptable based on the strengths of  Mg\,$b$ triplet
in their observed spectra.  Hence, the plausible reason
for the weak/absent MgH bands in the observed spectra of these stars
is a relatively lower abundance of hydrogen in their
atmospheres.  These giants may belong to the group
of helium enriched red giants of $\omega$\,Cen.
\end{abstract}


\section{Introduction}

A rare group of hydrogen-deficient (H-deficient)
and carbon rich supergiants in the order of 
their increasing effective temperatures are:
hydrogen deficient carbon (HdC) stars,
R\,Coronae Borealis (RCB) stars, and extreme helium (EHe) stars
\citep{hema12, pandey06a}.
The origin and evolution of these peculiar stars is not yet clear. 

The distances are not accurately known to any
of the Galactic H-deficient stars.
The position of a star on the HR-diagram
gives us an idea about its evolution and possibly its origin.
To place these stars on the HR-diagram, 
a survey was conducted to identify new H-deficient 
stars in the  most massive and brightest 
Galactic globular cluster (GGC): $\omega$\,Centauri.

The observed large spread in the metallicity ([Fe/H])
and the other abundance anomalies of $\omega$\,Cen cluster 
stars \citep{johnson10, marino11, simpson13, sollima05} including 
the existence of multiple stellar populations, the He-normal 
and He-enhanced or H-poor \citep{piotto05}, makes it an   
enigmatic GGC. 
The recent spectroscopic studies of \citet{dupree13} and \citet{marino14} 
confirms the existence of the He-enhanced stars in  
$\omega$\,Cen and NGC 2808, respectively. 
Hence, our survey also explores the H-deficiency or the 
He-enhancement in the sample giants of $\omega$\,Cen.

\section{Sample Selection and Observations}

Our survey is based on the Str\"{o}mgren photometric studies
of red giant stars in $\omega$\,Cen
by \citet{calamida09}.
Stars in the GGCs are expected to be 
homogeneous in metallicities ([Fe/H]).
But, there are a few  GGCs
which show a large dispersion in their metallicities.
One of them being the GGC  $\omega$\,Cen, 
with a range in [Fe/H]: 
$-$2.5 $<$ [Fe/H] $<$ $+$0.5 (\citet{johnson10} and 
references therein).
This spread in metallicity, which is not as 
expected for a GGC, is possibly an indication
for the presence of H-deficient stars in $\omega$\,Cen.
Note that, the metal lines are much stronger in the spectra
of  the H-deficient supergiants: the RCB and the HdC 
stars, when compared with those observed in the 
spectra of F- and G-type normal supergiants. 
This is attributed to the lower opacity in the 
atmosphere due to H-deficiency.
Hence, H-deficient stars appear metal richer than they 
actually are \citep{sumangala11}. Our
suspicion was that the metal rich giants of $\omega$\,Cen may
possibly be H-poor. In this survey the priority was given
to the giants in the metallicity range: $-$0.5 $>$ [Fe/H] $<$ $+$0.5.
However, for the sake of the completeness,
irrespective of the metallicity, all the giants,
brighter than 14.5 y magnitude (Str\"{o}mgren visual)
were considered for our study.
The other criterion that was applied to increase the
probability of finding H-deficient stars in our sample
was the (J-H)$_{0}$ and  (H-K)$_{0}$ colours (IR-colours).
The RCB and the HdC stars' distribution in the
IR-colours plot is distinct from the normal dwarfs and giants
\citep{feast97, Tisserand09}.  
The J, H and K magnitudes and the corresponding 
Galactic dust reddening for extinction correction were adopted from 
Two Micron All Sky Survey (2MASS)
Catalogue\footnote{http://irsa.ipac.caltech.edu/Missions/2mass.html}
and NASA/IPAC Infrared Science
Archive\footnote{http://irsa.ipac.caltech.edu/applications/DUST/}, 
respectively.

Note that, for observations, all the metal rich
giants (+0.5 $>$ [Fe/H] $>$ $-$0.5) were selected, irrespective
of their IR-colours -- these were 130 in number.
However, the metal poor giants ($-$0.5 $>$ [Fe/H] $>$ $-$2.5)
selected for observations, with  IR-colours like RCB stars,
were 40 in number.
Though, the sample of red giants from the core of
$\omega$\,Cen were not included in our sample
(to avoid confusion in identifying the giants in the crowded field),
many of the giants in the periphery
were double or multiple objects. The giants
which were not clearly resolved were excluded
from our observations.
Hence, only 34 of the 130 metal rich stars
and about 11 of the 40 metal poor stars were selected for
observations.

Low-resolution optical spectra for these selected red giants
of $\omega$\,Cen  were obtained from
the 2.34 m Vainu Bappu Telescope (VBT), 
Vainu Bappu Observatory,  equipped with the
Optomechanics Research spectrograph
\citep{prabhu98} and 1K $\times$ 1K CCD camera.
These spectra obtained using 600 l/mm
grating centered at H$\alpha$ line at 6563\AA\  
 were at a resolution of about 8\AA.
The data reduction and analyses were carried out using the 
 IRAF\footnote{The IRAF software is distributed by the
National Optical Astronomy Observatories under contract
with the National Science Foundation.}
(Image Reduction and Analysis Facility) software package.

\section{Analyses and Results}

The observed spectra of all the program stars were 
continuum normalized. 
The region of the spectrum (having maximum flux) free of absorption
lines is treated as the continuum point, and a smooth
curve passing through these points  is
defined as the continuum.
The well defined continuum in the spectrum of the 
sample metal poor giant, and in the spectrum of 
Arcturus with very high signal-to-noise (S/N),
is used as a reference for judging the continuum for the sample  
metal rich stars in the wavelength window 4900 - 5400A
including the Mgb triplet and the complete MgH band.
The analyses of the observed spectra of the program stars
were carried out based on the strengths
of the blue degraded (0,0) MgH band extending from 5330 to 4950\AA,
with the band head at 5211\AA, and the Mg\,$b$ lines at
5167.32\AA, 5172.68\AA\ and 5183.60\AA.
Based on the strengths of these features in the observed spectra, 
three groups were identified in our sample: (i) the metal
rich giants with strong Mg\,$b$ lines and the MgH band,
(ii) the metal poor giants with weak Mg\,$b$ lines and
no MgH band, and (iii) the metal rich giants with strong
Mg\,$b$ lines, but no MgH band.
To analyze the strengths of  the MgH band in the observed 
spectra of sample stars, the stars  with 
similar (J-K)$_{0}$ colours  ($\Delta$(J-K)$_{0}$\,$\sim$\,$\pm$0.1), 
and y magnitudes  ($\Delta$y\,$\sim$\,$\pm$0.5), that 
represent the effective temperatures ($T_{\rm eff}$) and 
surface gravities (log\,$g$), respectively, were selected.
The spectra of stars having similar 
y magnitude and (J-K)$_{0}$ colours were then compared with each other. 
From this comparison four stars 
were identified having weaker or absent MgH band than expected.
Two stars: 178243 and 73170 are from the
first group showing the strong Mg\,$b$ lines, but
weaker MgH band than expected for their stellar
parameters (see Figure 1). 
The other two stars: 262788 and 193804
are from the third group showing relatively  
strong Mg\,$b$ lines, but absent MgH band 
not as expected for their stellar parameters\footnote{ Note 
that, the contribution of the MgH lines to the Mg\,$b$ 
line strength makes the latter appear stronger in the 
first group than the third group.}
 (see Figure 2).
Judging by the observed strengths of the Mg\,$b$ lines
and the presence/absence of the MgH band expected for their
stellar parameters, the spectra of the giants 
178243, 73170, 262788 and 193804, suggest 
that their atmospheres are relatively 
H-poor. Hence, to confirm this suggestion,
the observed strengths of the MgH bands were
further analyzed by synthesizing the spectra
of these four stars along with the program
stars of first and third groups
for their adopted stellar parameters.
Note that, observed spectra with S/N $>$ 60 were analyzed.

A literature survey was done for the high-resolution spectroscopic
studies of the program stars. Many of the program stars were
found in  \citet{johnson10}.
The stellar parameters: $T_{\rm eff}$, log\,$g$,
and [Fe/H] for the program stars were
adopted from  \citet{johnson10}.
The  $T_{\rm eff}$ for the program stars were also 
determined using their photometric colours: (J-H)$_{0}$,
(J-K)$_{0}$, and (b-y)$_{0}$ using \citet{alonso99}'s 
empirical calibrations.
The log\,$g$ for the program stars were determined
using  the standard relation.
These determined $T_{\rm eff}$ and log\,$g$ values are 
within an agreement of $\pm$200K, and $\pm$0.2 (cgs units), and 
with a mean difference of $\pm$100K, and $\pm$0.1 (cgs units), respectively,
with those determined by \citet{johnson10}.
Hence, we have used our estimates of the
$T_{\rm eff}$ and log\,$g$ for the stars not
available in \citet{johnson10}.

The spectra were synthesized from 5100\AA-5200\AA\
that includes the Mg\,$b$ lines and
the (0, 0) MgH band.
For synthesizing the spectra, the atomic
lines were compiled from the 
standard atomic data sources, and all the 
atomic lines identified by \citet{hinkle00} were included.  
The (0, 0) MgH molecular line list was adopted from \citet{hinkle13}.
Synthetic spectra were generated by combining the LTE spectral
line analysis/synthesis code MOOG \citep{sneden73},
and the ATLAS9 \citep{kurucz98} plane parallel, line-blanketed
LTE model atmospheres with convective overshoot.
Spectrum of Arcturus, a typical red giant, was synthesized to
validate the adopted $gf$-values of the atomic/molecular lines.
Using the spectrum synthesis code, $synth$
in MOOG, the high resolution optical spectrum of Arcturus
was synthesized for the stellar parameters and
the abundances given by \citet{ramirez11}.
The synthesized spectrum was convolved with a
Gaussian profile with a width that represents
the broadening due to macroturbulence and the
instrumental profile.
Minimal adjustments were made to the abundances
of the atomic lines to obtain the best fit
to the observed high-resolution optical
spectrum of Arcturus \citep{hinkle00}.
The changes in the log $gf$ values were not more than 0.1 dex. 
A reasonably good fit was obtained to the MgH molecular lines
for the adopted isotopic values from \citet{mcwilliam88}.
The synthesized high-resolution spectrum was further convolved
by a Gaussian profile of width of about 8\AA,
to match with the observed low-resolution
Arcturus spectrum obtained from VBT.

A good match of the synthesized Arcturus spectrum to the
observed, both high- and low-resolution spectra, validates
the adopted line list for the adopted stellar parameters
of Arcturus. These checks on published analysis of Arcturus
are taken as evidence that our
implementation of the code MOOG, 
the LTE models, and the adopted line list
were successful for the syntheses of the red giants' spectra.
Hence, the spectra of the program stars
were synthesized  following the above procedure. The
synthesized spectra, for their adopted stellar parameters
and abundances, were then compared with the observed spectra.
From the studies of \citet{norris95}, the average
[Mg/Fe]\footnote{[Mg/Fe] = (Mg/Fe)$_{*}$ $-$
(Mg/Fe)$_{\odot}$} for the 
red giants of $\omega$\,Cen is about
$+$0.4\,dex over a metallicity range:
[Fe/H]=$-$2.0 to $-$0.7. Hence, in our synthesis
the [Mg/Fe]=$+$0.4 dex was adopted initially.
Since the subordinate lines of MgH band at about 5167\AA\ are
blended with the saturated Mg\,$b$ lines,
the subordinate lines of MgH band
in the wavelength window 5120--5160\AA\
were given more weight in our synthesis.
The best fit of the spectrum synthesized for the
adopted stellar parameters to the observed
was obtained by adjusting the Mg abundance,
and therefore estimating the
Mg abundance\footnote{log $\epsilon(Mg)$ = log $(Mg/H)$ $+$ 12.0,
this convention is used throughout this study.}
for the program star (see Figure 3 for example).
Note that, the derived Mg abundances are in
excellent agreement with the two common
stars in \citet{norris95} study.
The adopted stellar parameters and the derived Mg 
abundances  for the program stars 
(first and third group) are given in Table 1.
For all the normal first and third group stars,
our derived Mg abundances (mean [Mg/Fe]\,$\sim$\,$+$0.3 dex)
for their adopted stellar parameters, were as expected
for the red giants of $\omega$\,Cen,
with just four exceptions. These four 
exceptions are: 73170 and 178243 from the first group,
and 262788 and 193804 from the third group, that were
identified with the weak/absent MgH bands in their 
observed spectra. 
The Mg abundances derived
for these four giants are much lower than
that expected  (for details see Hema 2014).

\section{Discussion}

The two stars of the first group with strong Mg\,$b$ lines
and weaker MgH band are 73170 and 178243. 
The MgH subordinate lines to the redward and the blueward
of the Mg\,$b$ lines are clearly weaker in the observed spectra
of 73170 and 178243, when compared with the spectra
of stars with similar stellar parameters.
Figure 1 shows the spectra of the first group stars in the
order of their increasing $T_{\rm eff}$ from bottom to top.
This comparison clearly shows that the weaker MgH
bands in the spectra of these two stars
are not as expected for their adopted stellar parameters.
If the weaker MgH band is not due to the star's
$T_{\rm eff}$, log $g$ and [Fe/H], then the reason would be
a lower Mg abundance.
These stars are metal rich with strong Mg\,$b$ lines
in their observed spectra, indicating that the Mg abundance
in their atmospheres is as expected for their
metallicities. Hence, neither the
stellar parameters nor the Mg abundances are  possible
reasons for the weaker MgH bands observed in these stars. 
The only possible reason for the weaker MgH bands would be
a lower hydrogen abundance in their atmospheres.

For the  derived Mg abundance of 73170, 
the [Mg/Fe] is about $-$0.2. This value is
about $+$0.6 dex lower than the average,
and about $+$0.4 dex lower than the minimum
[Mg/Fe], derived for the $\omega$\,Cen giants
at the star's metallicity of [Fe/H]=$-$0.65, 
as reported by \citet{norris95}.
For the derived Mg abundance of 178243, the [Mg/Fe] is about $-$0.4.
 This value is about $+$0.8 dex lower than the
average, and about $+$0.6 dex lower than
the minimum  [Mg/Fe] derived  for the
red giants of $\omega$\,Cen at the
star's metallicity of [Fe/H]=$-$0.8 \citep{norris95}.
Nevertheless,  going by the observed strengths of the Mg\,$b$ lines and
the expected Mg abundance for the stars' metallicity, our
derived low Mg abundances for these two stars are unacceptable.
Hence, we emphasize that the weaker MgH bands in these two stars
are not due to the stellar parameters and the Mg abundances, but
most probably due to a relatively lower abundance of hydrogen
in their atmospheres.
The typical errors of about $\pm$100K on $T_{\rm eff}$,
determined from the high-resolution spectroscopic studies 
of red giants, were adopted. Note that, the syntheses of the MgH bands 
for the upper limit of the $T_{\rm eff}$ for these two stars
do not provide a fit to the observed spectrum (see Figure 3 for example).

The two third group stars, 262788 and 193804, have
similar stellar parameters, as given by \citet{johnson10}, and
their observed spectra are near identical.
Hence, treated as twins.
The observed spectra of these twins were
compared with the observed spectra of the first and 
second group stars of similar stellar 
parameters (see Figure 2).
The first group star, 251701, with similar stellar
parameters as the twins of the third group,
shows  clear presence of the
MgH band in the observed spectrum, but note
the absence of MgH bands in the observed spectra of
the twins. 
The reason for the absence of the MgH band
in the spectra of these twins would be
a lower Mg abundance in their atmospheres. However,
the Mg\,$b$ lines in their
observed spectra are strong, as expected for their metallicity
 and similar in strengths to those of the first group stars.
The Mg\,$b$ lines in the spectra of these twins are
fairly stronger than in the spectra of the
second group stars. Note that, the second group stars
are metal poor and have weak or no Mg\,$b$ lines (see Figure 2).
The strong Mg\,$b$ lines in the spectra of 
262788 and  193804, clearly indicate
that the Mg abundance is normal or as expected for the stars'
metallicity.  Hence, the stellar parameters and the lower Mg
abundances are ruled out as the possible reasons for the
absence of the MgH band in the spectra of these twins. The only
possibility could be a relatively lower abundance of hydrogen
in their atmospheres.

For 262788, the derived [Mg/Fe]$\leq$$-$0.6.
This is about $+$1.0 dex lower than
the average, or about $+$0.8 dex
lower than the minimum [Mg/Fe] value
as derived by \citet{norris95},
for the star's metallicity, [Fe/H]=$-$1.0.
For 193804, the derived [Mg/Fe]$<$$-$0.1.
This value is about $+$0.5 dex
lower than the average, or about
$+$0.3 dex lower than the minimum [Mg/Fe] value,
as reported by \citet{norris95},
for the star's metallicity [Fe/H]=$-$1.0.
Hence, going by the observed strengths of the Mg\,$b$ lines, and
the expected  Mg abundance for the stars' metallicity, our
derived low Mg abundances for these two stars are unacceptable.
This rules out the effects of stellar parameters and the
lower Mg abundances, as the possible reasons for
the absence of the MgH band in the observed spectra of 262788 and 193804.
The only possible reason for the
absence of MgH band could be a relatively lower abundance of
hydrogen in their atmospheres. 
The spectra synthesized for the upper limit of the $T_{\rm eff}$
for the stars 262788 and 193804, do not provide 
a fit to the observed spectra.
The observed weak/absent MgH band
in 214247, of the third group,  
is as expected for the star's warmer
$T_{\rm eff}$ and metallicity. The derived Mg 
abundance is as expected for the star's metallicity.

The weak/absent MgH band in the observed
spectra of these four giants inspite of
the presence of strong Mg $b$ lines,
may not be due to the uncertainty in adopted 
stellar parameters or
a lower Mg abundance. The only plausible reason
is a relatively lower abundance of hydrogen in their atmospheres.
Hence, we report the discovery of
four giants with relatively lower abundance of hydrogen
in their atmospheres.
These giants may belong to the group
of helium enriched
($n_{He}$/$n_{H}$ $\sim$ 0.16\,--\,0.2)\footnote{Note
that, the expected $n_{He}$/$n_{H}$ ratio and Y for the normal giants
are about 0.1 and 0.25, respectively.}
red giants of $\omega$\,Cen similar to those
found in the studies of \citet{dupree11}, for which
the  blue main sequence (bMS) stars may
be the progenitors. The double main sequence,
the red main sequence (rMS) and the blue main sequence (bMS),
in $\omega$\,Cen was  discovered
by \citet{anderson97}. 
The bMS stars differ from rMS stars by their
helium enrichment upto Y $\sim$ 0.38 with
the range: 0.35 $<$ Y $<$ 0.4 \citep{norris04, piotto05}.
The Y value for rMS stars is about 0.25 \citep{norris04, piotto05}.
From the studies of \citet{marino11}, unlike the observed 
Na-O anticorrelation for the metal poor giants of $\omega$\,Cen, 
the number of giants with Na-enrichment increases 
with the metallicity ([Fe/H]). 
%From spectroscopic studies on the red giants of 
%GGCs, \citet{bragaglia10} have noticed that
%the metal rich stars that show Na-enrichment, also show He-enrichment. 
Note that 11 stars of our metal rich sample, including the reported
4 H-poor or He-rich stars, show Na-enhancement \citep{marino11, johnson10}. 
One star, however, is Na-normal.
For rest of our metal rich sample of 5 stars, Na abundances are not available.
However, no obvious trend is seen in Al-Mg, as the Al-Mg anticorrelation
provides further clue for the He-enrichment in the metal 
rich giants \citep{ercole10}.

None of these newly discovered H-deficient giants show
the carbon features as seen in the
spectra of RCB stars and they do not exhibit RCB IR-colours.
Hence, these are not
the H-deficient stars of RCB type.
To ascertain a range in H-deficiency or
the helium enrichment of these newly discovered 
giants and the analyzed metal rich sample, it is essential to study these
stars by obtaining their high resolution spectra.

\section{Acknowledgements}

We thank the referee for a nice and constructive report. 
We thank David L. Lambert, H. C. Bhatt, and 
B. E. Reddy for fruitful discussions. 
We thank the VBO staff for their assistance.

%\bibliographystyle{apj}
%\bibliography{ref1}

%\end{document}

\begin{figure}
\centerline{\includegraphics[height=6.5in, keepaspectratio=true]{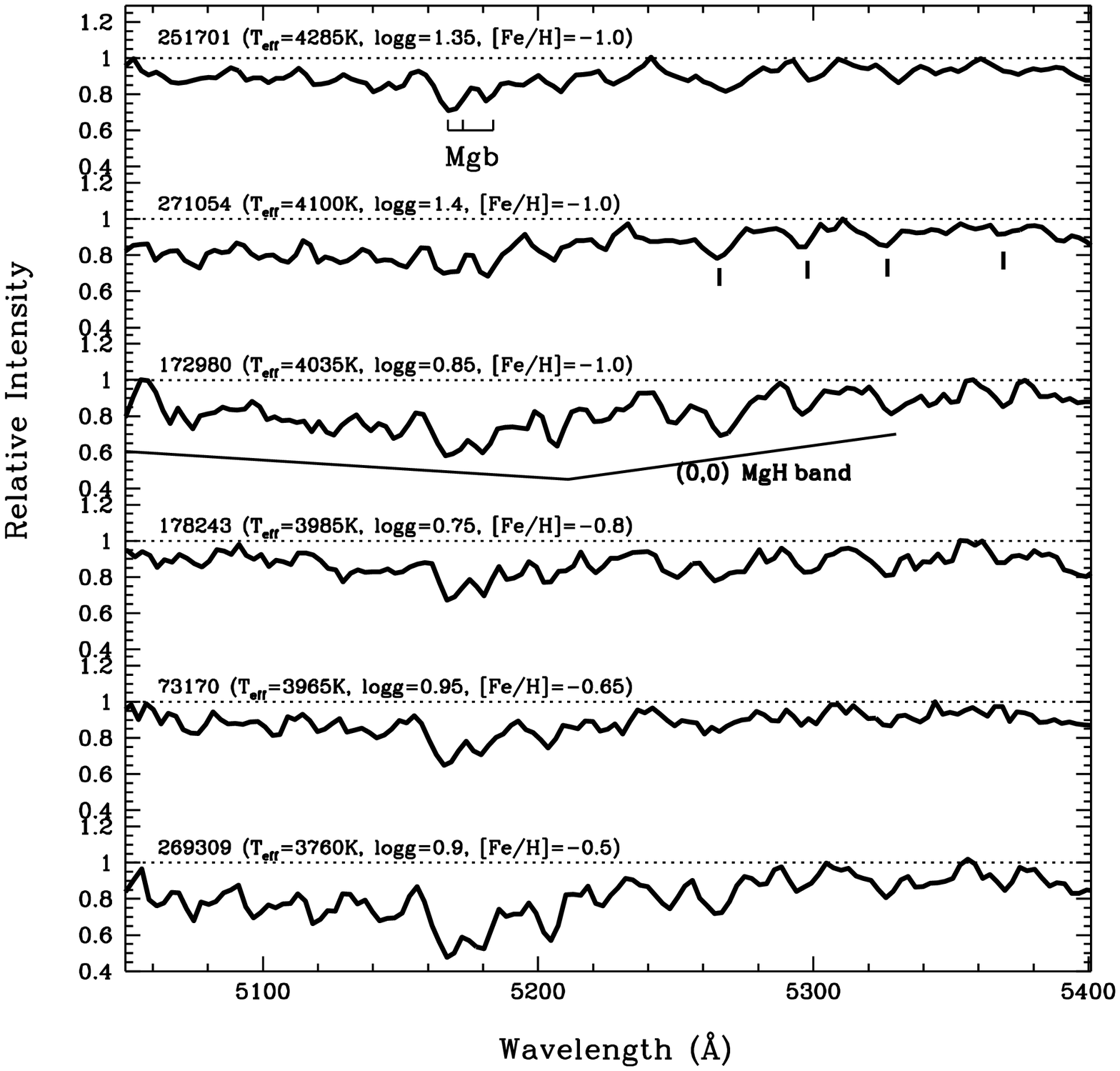}}
\caption{The spectra of the first group stars
are shown in the order of their increasing
$T_{\rm eff}$ from bottom to top. The strong Mg\,$b$ lines
and the (0, 0) MgH band are marked. The vertical lines marked
to the red of the Mg\,$b$ lines are Fe\,{\sc i} lines. }
\label{Figure}
\end{figure}

\begin{center}
\begin{deluxetable}{lcccclcrr}
\tablecolumns{9}
\tablewidth{0pc}
\tablecaption{The  stellar parameters, metallicities and
derived Mg abundances for the program stars
in the order of their increasing $T_{\rm eff}$.}
\tablehead{
\colhead{Star} & \colhead{Star(LEID)\tablenotemark{a}} & \colhead{S/N} &  \colhead{$T_{\rm eff}$}   & \colhead{log\,$g$} & \colhead{[Fe/H]} & \colhead{Group} & \colhead{log $\epsilon$(Mg)} & \colhead{[Mg/Fe]}}
\startdata
269309 &   \nodata   & 70 & 3760 & 0.90 & $-$0.5 & First & 7.1$\pm$0.2 & 0.0 \\
73170  & 39048 & 100 & 3965 & 0.95 & $-$0.65 & First & 6.75$\pm$0.2 & $-$0.2\\
178243 & 60073 & 100 & 3985 & 0.75 & $-$0.8 & First & 6.4$\pm$0.2 & $-$0.4\\
172980\tablenotemark{b} & 61067 & 110 & 4035 & 0.85 & $-$1.0 & First & 7.0$\pm$0.2 & $+$0.4\\
178691 & 50193 & 110 & 4075 & 0.65 & $-$1.2 & First & 6.6$\pm$0.2 & $+$0.2\\
271054 &   \nodata   & 100 & 4100 & 1.40 & $-$1.0 & First & 6.7$\pm$0.2 & $+$0.1\\
40867  & 54022 & 110 & 4135 & 1.15 & $-$0.5 & First & 7.2$\pm$0.2 & $+$0.1\\
250000 &   \nodata   & 90 & 4175 & 1.40 & $-$1.0 & First & 6.9$\pm$0.2 & $+$0.3\\
131105 & 51074 & 80 & 4180 & 1.05 & $-$1.1  & First & 6.9$\pm$0.2 & $+$0.4\\
166240\tablenotemark{b} & 55101 & 60 & 4240 & 1.15 & $-$1.0 & First & 6.8$\pm$0.2 & $+$0.2\\
262788 & 34225 & 110 & 4265 & 1.30 & $-$1.0 & Third & $<$6.0$\pm$0.2 & $<-$0.6\\
251701 & 32169 & 100 & 4285 & 1.35 & $-$1.0 & First & 7.0$\pm$0.2 & $+$0.4\\
193804 & 35201 & 80 & 4335 & 1.10 & $-$1.0 & Third & $<$6.5$\pm$0.2 & $<-$0.1\\
5001638 &   \nodata   & 150 & 4400 & 1.6 & $-$0.5 & First & 7.3$\pm$0.2 & $+$0.2\\
270931 &   \nodata   & 100 & 4420 & 1.25 & $-$0.5 & First & 7.2$\pm$0.2 & $+$0.1\\
214247 & 37275 & 60 & 4430 & 1.45 & $-$1.5 & Third & 6.5$\pm$0.2 & $+$0.4\\
216815 & 43475 & 80 & 4500 & 1.85 & $-$0.6 & First & 7.3$\pm$0.2 & $+$0.3\\
14943 &  53012 & 100 &  4605 &  1.35 & $-$1.8 & Second & $<$6.7$\pm$0.2 &  $<+$0.9 \\
\enddata
\tablenotetext{a}{Stellar parameters are from \citet{johnson10} 
for the giants with LEID identification.}
\tablenotetext{b}{Common stars with the sample of \citet{norris95}.
Norris \& Da Costa report [Mg/Fe] of 0.53 and 0.27 for
172980 and 166240, respectively.}
\end{deluxetable}
\end{center}

\begin{figure}
\centerline{\includegraphics[height=6.5in, keepaspectratio=true]{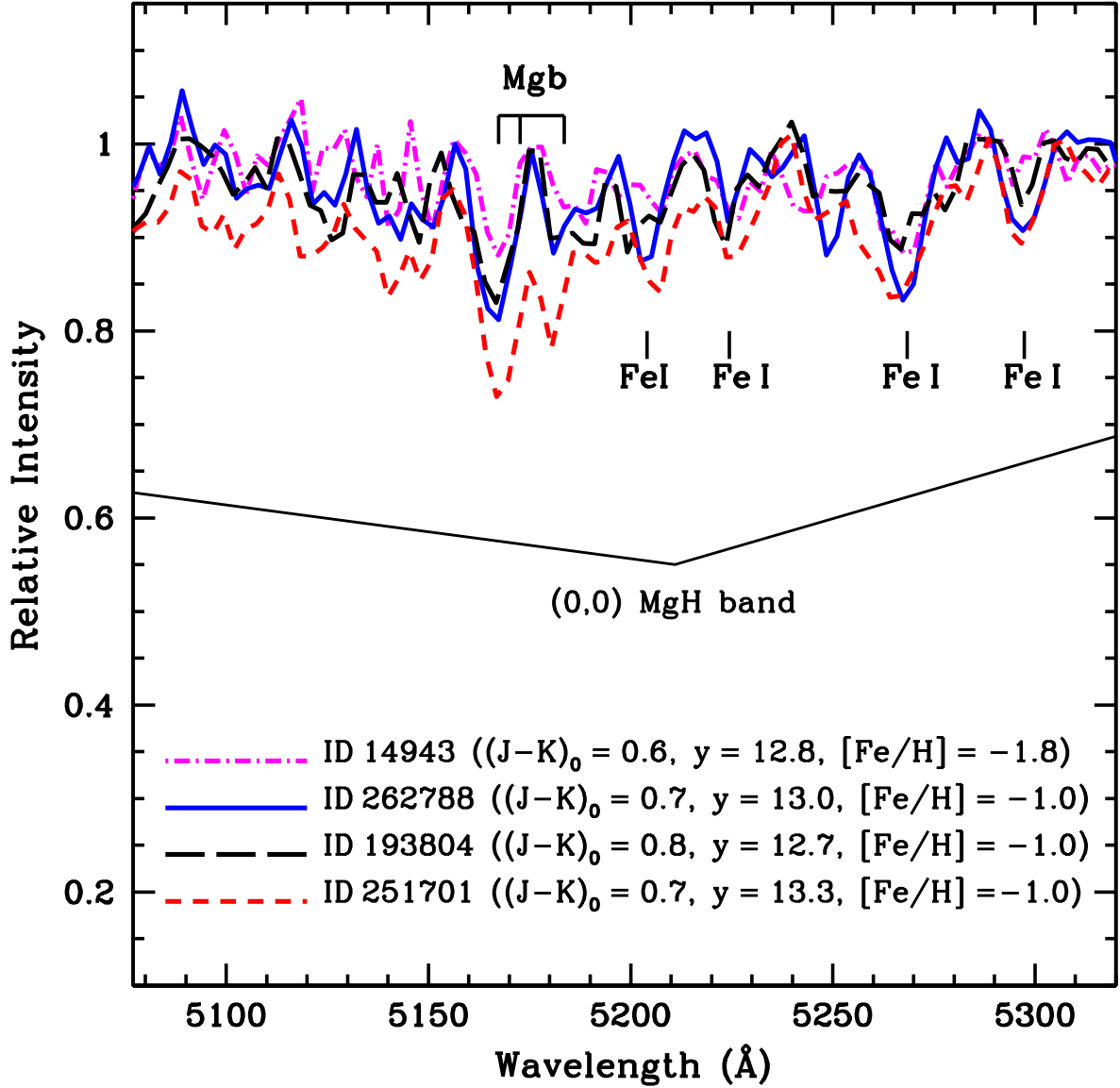}}
\caption{Figure shows the observed spectrum
of 262788  and 193804,
the third group stars-twins,
compared with the observed spectrum of 251701,
the first group star.
Also shown is the observed spectrum of 14943,
the second group star.
The key features such as Mg\,$b$ lines, the MgH band and
the Fe\,{\sc i} lines are marked.}
\label{Figure}
\end{figure}

\begin{figure}
\centerline{\includegraphics[height=6.5in, keepaspectratio=true]{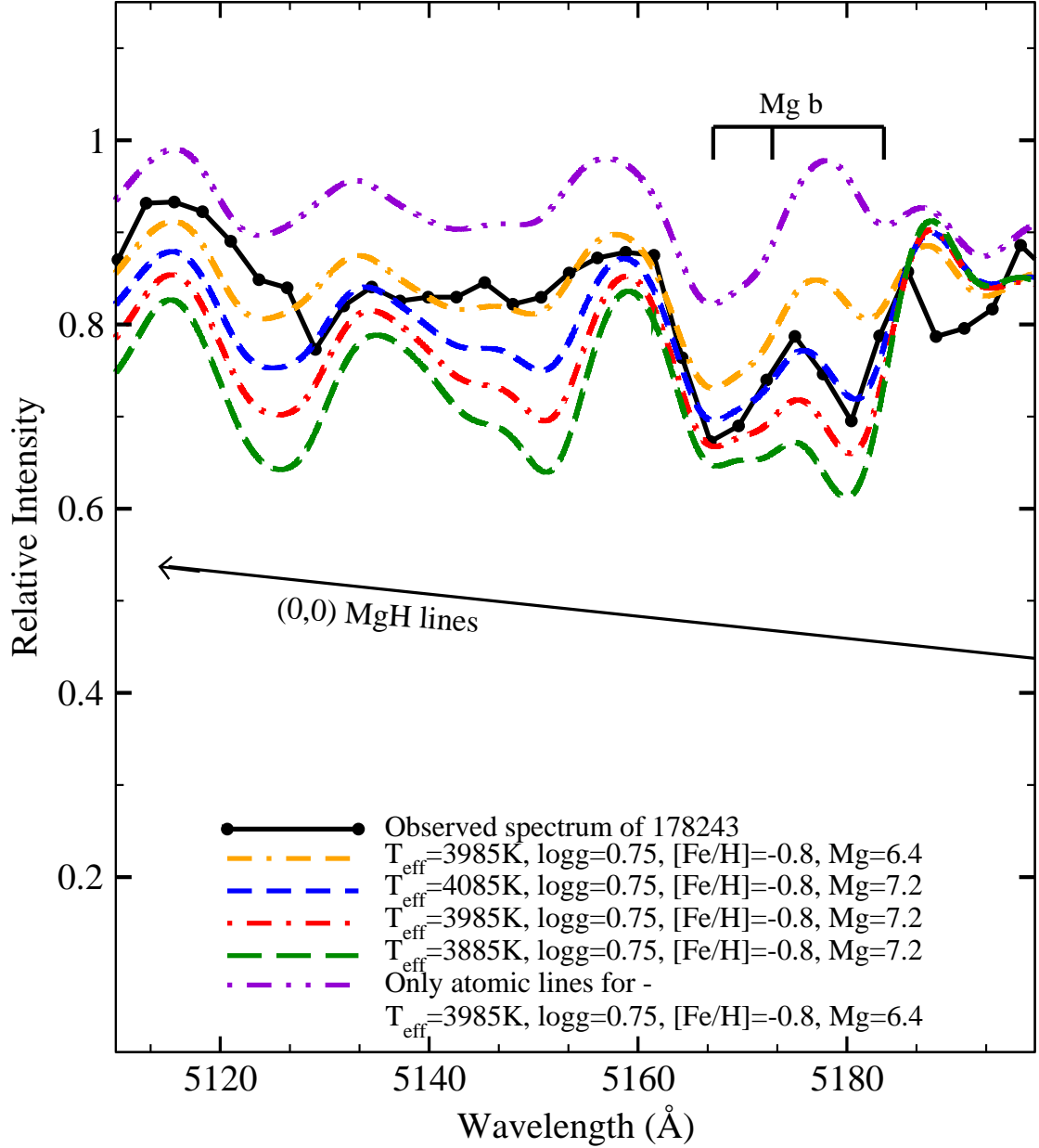}}
\caption{Figure shows the observed spectrum and the synthesized spectra
for the star 178243. The spectrum synthesized for the best fit 
value of [Mg/Fe], and for different $T_{\rm eff}$ for the 
expected [Mg/Fe] of $\omega$ Cen red giants are shown in the 
figure -- see key on the figure.}
\label{Figure}
\end{figure}

\end{document}